\documentclass[draft]{agujournal}

\journalname{Geophysical Research Letter}

\begin{document}

%
%


\title{Sensitivity of the Geomagnetic Octupole to a Stably Stratified Layer in the Earth's Core}

%
%




\authors{C. Yan\affil{1} and S. Stanley\affil{1,2}}


\affiliation{1}{Department of Earth and Planetary Science, Johns Hopkins University, Baltimore, MD 21218, USA.}
\affiliation{2}{Johns Hopkins University, Applied Physics Laboratory, Laurel, MD 20732, USA.}




\correspondingauthor{Chi Yan}{cyan10@jhu.edu}




\begin{keypoints}
\item Most standard Earth-like numerical dynamo simulations cannot reproduce the geomagnetic octupolar component from the past 10,000 years.
\item We implement a stably-stratified layer at the top of the core in dynamo models to determine the effects on the magnetic octupole.
\item We find that specific ranges of stable layer thickness and stability are needed to match geomagnetic observations.
\end{keypoints}

%
%


\begin{abstract}
Current ``Earth-like" numerical dynamo simulations are able to reproduce many characteristics of the observed geomagnetic field. One notable exception is the geomagnetic octupolar component. Here we investigate whether a stably stratified layer at the top of the core, a missing ingredient in standard dynamo simulations, can explain the observed geomagnetic octupole. Through numerical simulations, we find that the existence of a stable layer has significant influence on the octupolar-to-dipolar ratio of the magnetic field. Particularly, we find that a 60 km stable layer with relatively strong stability or a 130 km layer with relatively weak stability are compatible with the observations, but a 350 km stable layer, as suggested by recent seismological evidence, is not compatible with Earth's octupole field over the past 10,000 years. 
\end{abstract}

%
%

%


%
%
%
%
\section{Introduction}
The geomagnetic field is generated through dynamo action operating in Earth's liquid outer core, where convection is driven by thermal and compositional buoyancy forces as Earth slowly cools and the inner core solidifies. Archeomagnetic and paleomagnetic data demonstrate that Earth's field is axially-dipolar dominated on long timescales and exhibits variability on various timescales including westward drift, excursions and aperiodic reversals. Numerical simulations of dynamo action are used to investigate the mechanism responsible for generating Earth's magnetic field. Comparing the results of these simulations to observations of the present and past geomagnetic field provides vital information on processes occurring in Earth's deep interior. Geodynamo models aim to reproduce the salient features of the geomagnetic field including the dipole dominance, spatial power spectrum and temporal characteristics of the variability. Researchers have proposed quantitative criteria for determining whether a simulated field is ``Earth-like" and determined regions of parameter space where such fields occur \citep[][]{C2010, D2014}. Models in this parameter space are believed to provide the best insights into Earth's dynamo processes.

However, one large-scale field characteristic that many Earth-like models cannot reproduce is the octupolar component of the magnetic field (Figure 1). Here we use Gauss coefficients to represent the different modes of the magnetic field morphology. Outside the fluid core, where the magnetic field $\mathbf{B}$ can be represented as the gradient of a scalar potential $V$, the Gauss coefficients are defined by the expression
 \begin{equation}
  V(r,\theta,\phi) = r_e \sum^\infty_{l=1} \sum^l_{m=0} [g^m_l cos m\phi + h^m_l sin m\phi] {(\frac{r_e}{r})}^{l+1} P^m_l(cos\theta)
 \end{equation}
where $r$ is radius, $\theta$ is co-latitude, $\phi$ is longitude, $r_e$ is the radius of Earth's surface, $g^m_l$ and $h^m_l$ are Gauss coefficients, $l$ and $m$ are spherical harmonic degree and order, respectively, and $P^m_l(cos\theta)$ are associated Legendre polynomials. The three largest zonal signals are the dipole ($g^0_1$), quadrupole ($g^0_2$) and octupole ($g^0_3$). Following previous studies, we scale the octupole and quadrupole components to the dipole component ($g^0_3/g^0_1$ and $g^0_2/g^0_1$) for comparison.
A standard Earth-like model tends to produce an octupolar-to-dipolar ratio ($g^0_3/g^0_1$) that is always positive and larger than observational values over the past 10,000 years. This discrepancy between a standard model and the paleomagnetic model suggests that another ingredient may be necessary in the models to properly simulate Earth's dynamo processes. 

\begin{figure}[h]
\centering
\includegraphics[width=25pc]{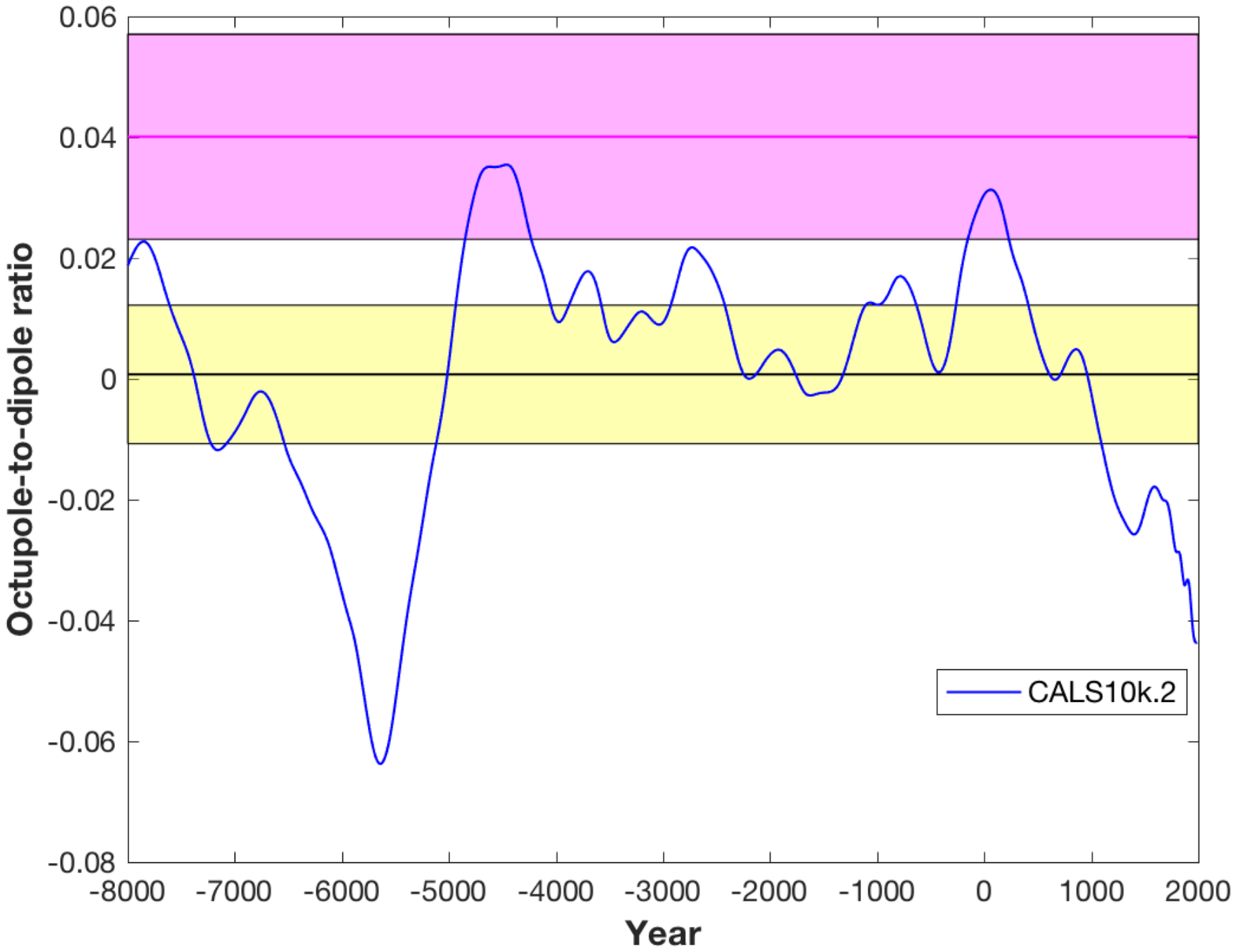}
\caption{The magnetic octupole-to-dipole ratio for paleomagnetic observations from the past 10,000 years from CALS10k.2 \citep{Con2016} and for an Earth-like dynamo model that does not include a stable layer (Model 0 of Table 1). The simulation time is re-dimensionalized through the magnetic diffusion time scale $d^2/\eta$ to have the same time span as in the CALS10k.2 model. The magenta line gives the time-averaged value from the model and the shaded magenta region represents the standard deviation about the average. The observational values are shown in the solid blue line with the black line representing the average and the shaded yellow region representing the standard deviation about the average.}
\end{figure}
 
Here we investigate whether a stably stratified layer at the top of Earth's core could be that ingredient. The presence of such a layer has long been proposed based on evidence from seismology \citep{tanaka93} and geomagnetic secular variation \citep[][]{b1993,b1999}. However, there is disagreement on the thickness and stability of this layer; properties which depend on the stable layer's origins. Recent seismological evidence suggests the layer may be over 300 km thick \citep[][]{tang15, Ka2018}) whereas secular variation studies suggest a thickness between $60 - 140$ km \citep{G2007, B2014}). Proposed origins for such a layer include a sub-adiabatic temperature gradient in the upper core due to its high thermal conductivity, producing a stable layer with thickness that may range from  $\sim 100$ km \citep{li1998} to $\sim 740$ km \citep{G2015}; compositional layering due to light element expulsion from inner core crystallization, resulting in a $\sim 250$ km thick layer \citep{H2013}; barodiffusion in the core, with a $\sim 100$ km thick layer \citep{G2012}; or a relic of merging cores from giant impacts early in Earth's history, with a $\sim 300$ km thick layer \citep{L2016}.

In this study we use numerical dynamo simulations to evaluate the effects of a stably stratified layer at the top of Earth's outer core on the resulting geomagnetic octupole.  Details of the numerical methods can be found in Section 2, results in Section 3, discussions in Section 4, and conclusions in Section 5. 

\section{Numerical Methods}
 We use the numerical dynamo model mMoSST \citep{K2008} to solve the coupled equations governing dynamo action in a fluid, electrically conducting, rotating outer core surrounding a solid, electrically conducting inner core. This model has been shown to reproduce benchmark results \citep{C2001}. 
Further details on the relevant dynamo equations, non-dimensional parameters and the numerical method can be found in \citet{K99} and \citet{K2008}. We additionally implement a stably stratified layer at the top of the core, where the layer stability is maintained through the background co-density gradient, in a similar manner as previous dynamo studies \citep[e.g.][]{S2004, S2008, Ch2008}. Further details on the non-dimensional parameters, model equations and implementation of stratification can be found in the supplementary material.

Due to numerical constraints, dynamo simulations cannot operate with realistic Earth-like parameters. However, scaling laws can be used to determine combinations of computationally attainable parameters that can produce dynamo generation with Earth-like characteristics. \citet{C2010} proposed conditions for an Earth-like dynamo model by defining quantitative criteria evaluating the level of agreement of the output from a numerical simulation with observed properties of the geomagnetic field morphology. We adopt their $\chi^2$ criterion, which is composed of four separate quantities, to evaluate the performance of our models. These four quantities are: (1) the ratio of the power in the axial dipole component to the power in the rest of the magnetic field, (2) the ratio of the power in the equatorially antisymmetric and symmetric magnetic field, (3) the ratio of the power in zonal and non-zonal non-dipole magnetic field, and (4) the concentration factor of magnetic flux at the core surface. 

For this study, we consider model C1-4* from \citet{D2014}, which satisfies the $\chi^2$ criterion but does not reproduce the observed $g^0_3/g^0_1$ ratio (Figure 1). We add stable layers of different thicknesses, i.e. 60 km, 130 km and 350 km, and strength of stratification to this model to determine whether it is possible to match both the $\chi^2$ criterion and the $g^0_3/g^0_1$ ratio. We also examine a case with no stable layer for comparison. Numerical simulation details are listed in Table 1. We use the Boussinesq approximation and apply co-density boundary conditions of fixed buoyancy at the inner core boundary (ICB) and fixed buoyancy flux at the core mantle boundary (CMB); no-slip boundary conditions on the velocity field; and magnetic field boundary conditions at the ICB for a finite electrically conducting inner core with equal conductivity to the outer core and at the CMB for an insulating mantle. We use finite differencing in the radial direction with 58 Chebyshev collocation points. Each spherical shell is resolved in latitude and longitude using spherical harmonics with maximum degree and order $l_{max} = 31, m_{max} = 23$. Our models are resolved. For example, simulations with $l_{max} = 50$, $m_{max} = 41$ and 78 radial points produce similar power spectra from degree 1 to 31 and the power in the highest degree $l_{max} = 50$ is 12 orders of magnitude smaller than the power in the lower degree $l_{max} = 31$. \par

\begin{table}[h]
 \caption{Model parameters: $\Delta r$ is the thickness of the stable layer. $(N/2\Omega)^2$ measures the strength of the stable layer stratification where $N$ is the $Brunt-V\ddot{a}is\ddot{a}l\ddot{a}$ frequency and $\Omega$ is the angular velocity. Other non-dimensional parameters relevant to the models are held fixed at Ekman number of $ E = 1.2 \times 10^{-4}$ , Prandtl number $Pr= 1$, the magnetic Prandtl number $Pm = 2$ and the modified Rayleigh number $R_a = 8.33 \times 10^5$ (note that the definition of these nondimensional numbers is given in the supplementary information).
}
 \centering
 \begin{tabular}{lll}
 \hline
  Model  & $\Delta r$ ($km$) 	& $(\frac{N}{2\Omega})^2$  \\
 \hline
   0 & 0 &  0	 \\
 \hline
   1 & 60 &  [0 $\to$ 1.36] \\
 \hline
   2 & 130 &  [0 $\to$ 1.36] \\
 \hline
   3 & 350 &  [0 $\to$ 1.36] \\
 \hline
\end{tabular}
 \end{table}

\section{Results}
Figure 2 shows the $g^0_3/g^0_1$ ratio averaged over $10,000$ years for our models. Although we ran our simulations for longer than 10,000 years, we chosen a random 10,000 year window in our simulations to present results here and confirmed that other randomly-chosen 10,000-yr windows produced similar results. The historical $g^0_3/g^0_1$ ratio from paleomagnetic model CALS10k.2 is also shown in the yellow shaded region. For the control model with no stable layer, $g^0_3/g^0_1$ is constantly larger than the values in CALS10k.2 model and never produces the negative values seen in CALS10k.2 model. However, for models with a stable layer, the average $g^0_3/g^0_1$ ratio decreases as the layer stability increases for all values of stable layer thickness we investigated. The ratios tend towards an equilibrium value as $(N/2\Omega)^2$ increases. This suggests that once a layer has become stable enough to fully inhibit convective flows (see Figure S3), there is no further effect on the $g^0_3/g^0_1$ ratio. Examining the cases with different stable layer thicknesses demonstrates that the thicker the layer, the more $g^0_3/g^0_1$ is impacted, both in terms of average values and variations.
 
  \begin{figure}[h]
\centering
\includegraphics[width=25pc]{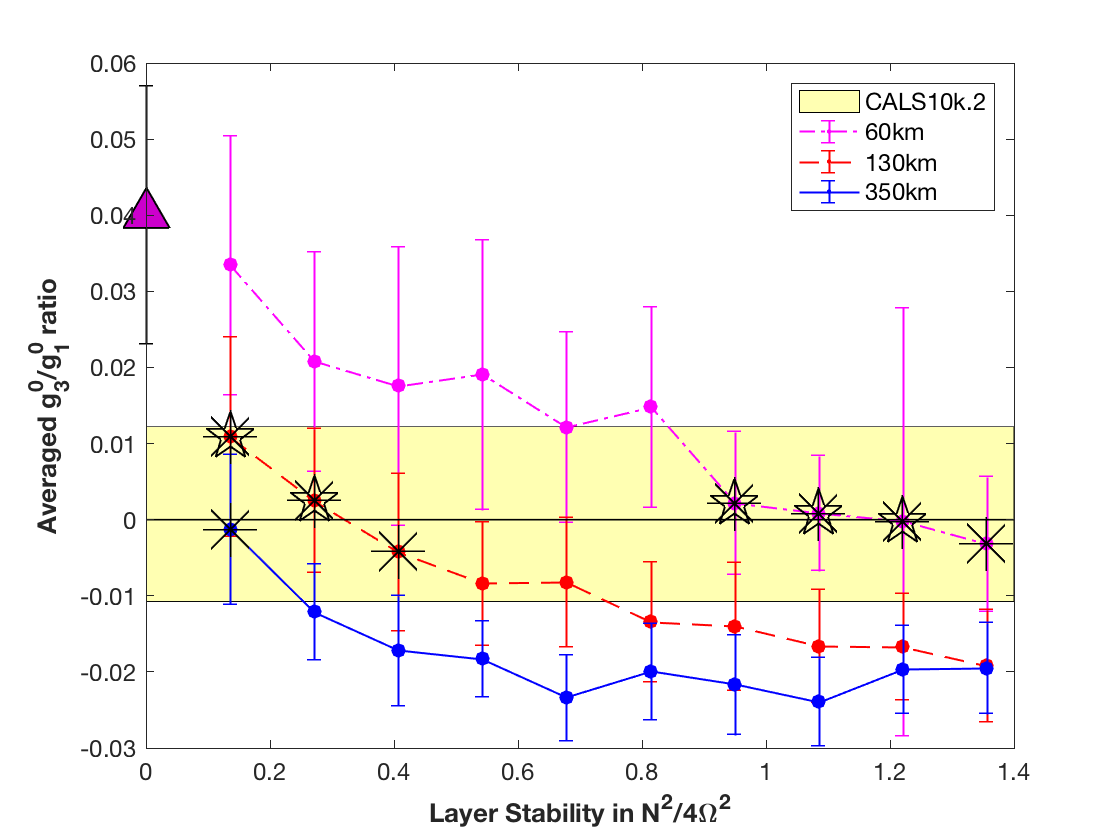}
\caption{The $g^0_3/g^0_1$ ratio as a function of stable layer properties. The error bars show the standard deviation in time of the $g^0_3/g^0_1$ ratios about the average values given by the circles. The model represented by a triangle is the control model 0 from Table 1 with no stable layer. Asterisks mark cases producing compatible $g^0_3/g^0_1$ ratios compared to CALS10k.2 and pentagrams mark cases that also meet the criterion $\chi^2$.}
 \end{figure}

 To determine which of our models can reproduce the most Earth-like characteristics, we first exclude models that don't produce the historical $g^0_3/g^0_1$ ratio, namely, we exclude models that don't produce time averages that are consistent with the observations as well as models that don't produce standard deviations that include both positive and negative values (e.g. the lack of positive $g^0_3/g^0_1$ values is why models with $\Delta r = 130$ km and $(N/2\Omega)^2 = 0.54$ and $0.68$ are excluded even though their mean values are consistent with the observations). This leaves 8 cases out of 33 simulations which are marked by asterisks in Figure 2. Next we exclude models that do not meet the quantitative criterion $\chi^2$ from \citet{C2010}, leaving 5 cases marked by pentagrams in Figure 2 that comply with standards for ``good" agreement ($\chi^2 < 4$) between the simulated field and the paleomagnetic model CALS10k.2 (Table S1) as well as meeting the $g^0_3/g^0_1$ ratio constraint. For visualizations of the radial magnetic field, see Figure S4(a). \par

We also investigated the zonal quadrupole to dipole ($g^0_2/g^0_1$) ratio in our simulations (Figure S1, S2). A standard Earth-like model without a stable layer is able to reproduce the historic values for this ratio and the addition of a stable layer in our models did not affect the ratio. Our results are therefore also consistent with observations for the quadrupolar field, although this wasn't an issue for the standard model to begin with.  The lack of dependence of the zonal quadrupole on the presence of a stable layer may be due to the fact that the zonal quadrupole belongs to a different dynamo symmetry family than the zonal octupole and dipole and is therefore generated by different convective modes. \par

Previous work has also demonstrated that a spatially variable CMB heat flux pattern can affect the zonal octupole component \citep[][]{B2000, He2013}.  For example, a surface spherical harmonic degree-2, order-0 ($Y^0_2$) pattern of heat flux perturbation at the CMB has been shown to result in a positive $g^0_3/g^0_1$ ratio that can better match paleomagnetic data from 250 Myr ago \citep{B2000}. It is thus necessary for us to examine the possible effects of the CMB heat flux variation over the past 10,000 years to disentangle the possible influences of CMB heat flux variation and the stable layer on the $g^0_3/g^0_1$ ratio. We have therefore imposed the current era's dominant heat flux signature ($Y^2_2$ pattern) discerned from mantle tomography on the CMB in our models to investigate the effects on the $g^0_3/g^0_1$ ratio. \par
 
  \begin{figure}[h]
\centering
\includegraphics[width=25pc]{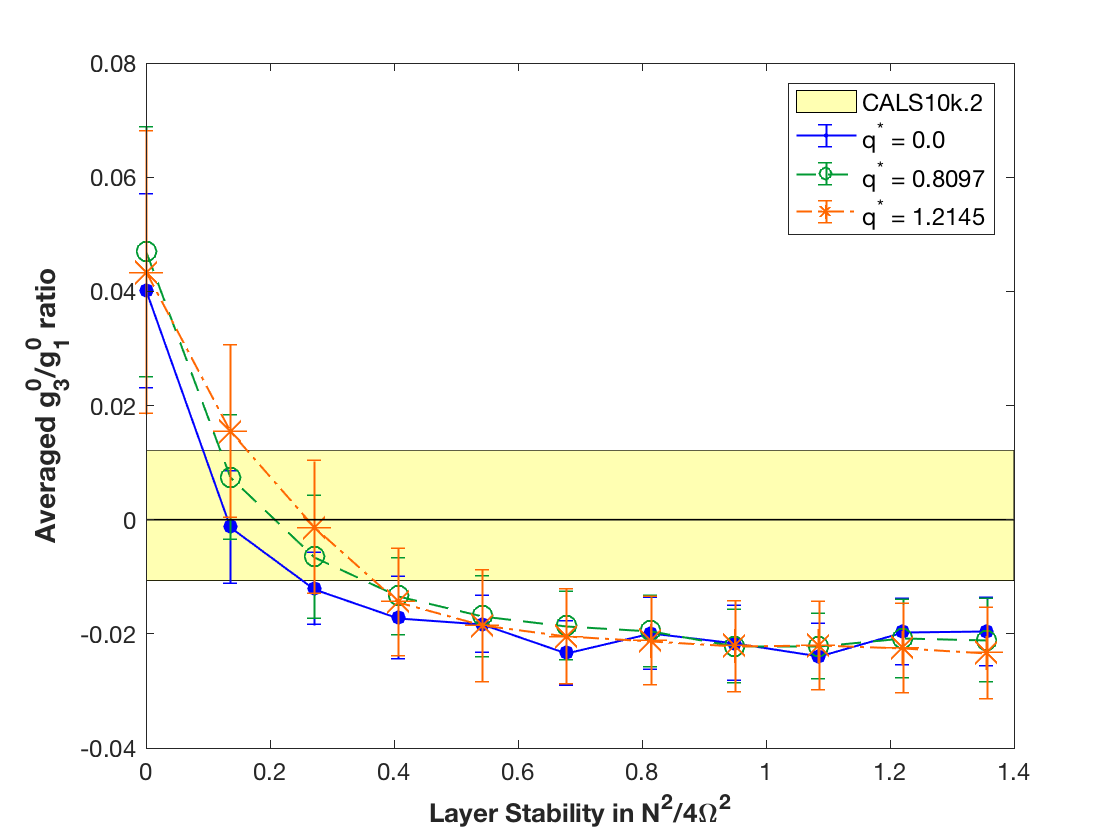}
\caption{The sensitivity of the ratio $g^0_3/g^0_1$ to the modern day heat flux variation pattern at the CMB, for a model with a 350 km stable layer. The blue line is the same as that in Figure 2. $q^*$ is the ratio of the heat flux anomaly divided by twice the average heat flux at the CMB.} 
\end{figure}

Figure 3 shows that: 1) when there is no stable layer, adding a $Y^2_2$ pattern of heat flux variation doesn't change the average  ratio although it causes a larger variation, as suggested previously by \citet{B2000}, and 2) a $Y^2_2$ pattern of heat flux variation did not significantly affect the $g^0_3/g^0_1$ ratio in our models. As a result, we conclude that the $g^0_3/g^0_1$ ratio is insensitive to the largest component of the modern era CMB heat flux pattern. \par

\section{Discussion}
Our study demonstrates that a stably stratified layer at the top of Earth's core may be necessary to explain the zonal octupolar component of the geomagnetic field over the past 10,000 years. 
It should be noted that other magnetic models such as CHAOS-4 \citep{O2014} and gufm1 \citep{J2000} have higher spatial and temporal resolutions compared to CALS10k.2 used in this study. However, those models only cover short time periods (recent decades) and therefore represent more of a snapshot of core processes making it unclear how representative they are of longer-term behavior. We wanted to compare average behavior on longer timescales and the CALS10k.2 model allowed for that. That being said, if we were instead to assume that the CHAOS-4 and gufm1 models from data over the past couple of decades was a better proxy of the average behavior of the large-scale components of the Earth's magnetic field over the past 10,000 years, then the prediction would be for an octupole-to-dipole ratio of $-0.0453\pm0.0002$ which is even further removed from the standard dynamo models without a stable layer than the data from the CALS10k.2 model. This would suggest that an even larger correction to the models would be needed (e.g. a thicker, more stable layer, or other new features in the model). We therefore feel we are making the conservative choice in this study by using the CALS10k.2 data as representative of average behavior over the past 10,000 years.

 Almost all models with a 350 km thick stable layer fail to match the geomagnetic octupolar constraint. The exception is our model with weak layer stability $(N/2\Omega)^2 = 0.14$ but this model fails to match the Earth-like $\chi^2$ criterion. This suggests discrepancy with recent seismic claims \citep{tang15, Ka2018} unless the seismic observations are capturing a physical process that is not being considered in our modeling approach of the stably stratified layer, (in particular, since the models operate in a parameter regime far from that of Earth's core.) \par

The reason for the stable layer's influence on the octupolar component of the magnetic field resides in the dynamo mechanism itself. Figure 4(a) shows that there is amplified power in velocity modes $(l,m) = (3, 0)$ and $(5, 0)$ due to the presence of a stable layer where models with larger layer stabilities lead to stronger amplification in these zonal flows. Dynamically as the stable layer is implemented, it forces thermal wind in the outer core (e.g. see Figure S4(b)) to be concentrated into the deeper region of the outer core, which results in a strengthened signature in the zonal octupolar toroidal kinetic energy. However, future studies are needed to scale the modeled zonal flows to these flows in real Earth conditions.\par
Figure 4(b) shows that models with a stable layer only see amplification in the $g^0_3/g^0_1$ ratio and not other magnetic modes. The amplified octupolar zonal flow can lead to creation of the magnetic octupole field through a two-step dynamo mechanism described with the \citet{B1954} formalism as:\\
			Step 1: ($T^0_3 S^1_2 T^1_1$)\\
			Step 2: ($S^1_2 T^1_1 S^0_3$)\\
Step 1 is described by an Adams-Gaunt integral which involves the octupolar zonal flow ($T^0_3$) acting on the poloidal magnetic field ($S^1_2$) to generate a toroidal magnetic field ($T^1_1$). Step 2 is described by an Elsasser integral where a poloidal flow ($S^1_2$) acts on the toroidal magnetic field ($T^1_1$) which was generated in Step 1 to generate the new magnetic zonal poloidal octupole field ($S^0_3$) that is observed. Similar to the above two-step dynamo mechanism, there are other paths leading to the creation of the magnetic octupole field through the amplified zonal flow in degree 5.
\par
\begin{figure}[h]
\centering
\includegraphics[width=25pc]{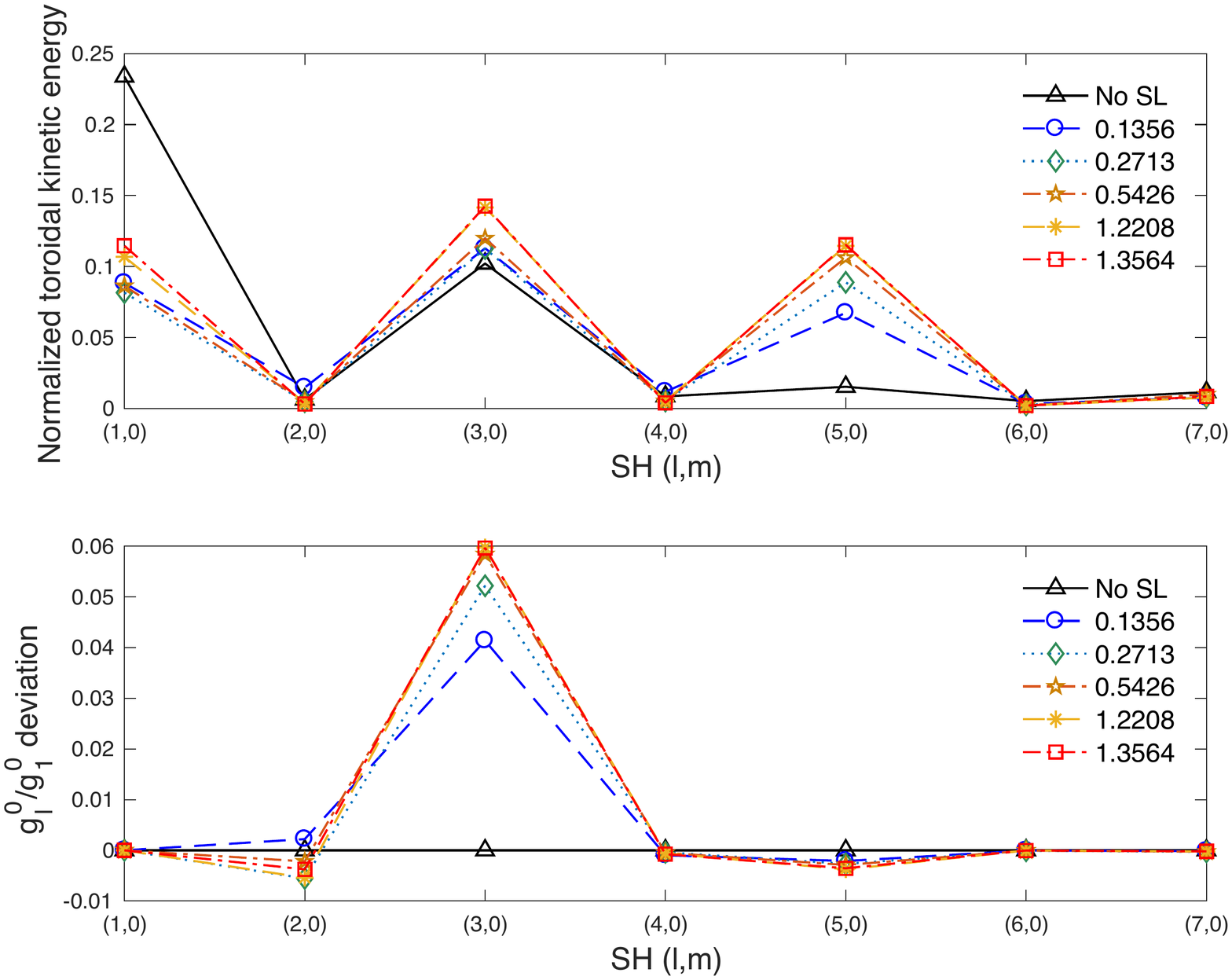}
\caption{(a) Normalized toroidal kinetic energy as a function of zonal spherical harmonic mode and (b) Deviation of the $g^0_l/g^0_1$ ratio in the 350 km case with different stratification $(N/2\Omega)^2$ values, averaged over 1 magnetic diffusion time. Both x axes list the sequential zonal spherical harmonic (SH) degree, the y axis in (a) shows the fraction of the toroidal kinetic energy of a single mode whereas the y axis in (b) shows the deviation of the $g^0_l/g^0_1$ in that model compared to the $g^0_l/g^0_1$ value without a stable layer (SL).} 
\end{figure} 
\par
A recent geodynamo study investigating partially stratified regions, where convection can occur locally in a stable layer due to strong thermal influences from the CMB, also found that the octupolar field is affected by the presence of a stable layer. However, they found the  $g^0_3/g^0_1$ ratio increased with layer stability and always produced a positive ratio \citep{O2017}. Their results would therefore not be consistent with observations, although it was not the purpose of that study to match observations. The reason for the differences between our studies is likely related to the fact that they were performed in different parameter regimes. This demonstrates that the effects we observe may be limited to the specific scenario appropriate to an Earth-like dynamo regime.\par

The model with 130 km stable layer and $(N/2\Omega)^2 = 0.27$ in this study is similar to the result in \citet{B2014} with a 140 km layer and $(N/2\Omega)^2 \approx 0.26$. However, the level of stratification investigated here is smaller than some previous predictions for what is expected in Earth's stable layer. For example, the \citet{G2012} barodiffusion model has $(N/2\Omega)^2$ ranging from 100 $\sim$ 400. However, the flattening of the $g^0_3/g^0_1$ ratios in our results (e.g. in Figure 2) suggests that further increasing the stratification in our models would not significantly affect the results. By this point, the layer is fully stratified with extremely small radial motions (see Figure S3). This suggests that there is no entrainment by the underlying convection into the stable region. \par 

It is worth noting that a stably stratified layer might not be the only mechanism affecting the magnetic octupole. For example,
\citet{B2000} found that the $Y^0_2$ CMB heat flux variation pattern from 250 Ma increases the $g^0_3/g^0_1$ ratio compared to a model with no CMB heat flux variations. 
Modern day values of $Y^0_2$ may be quite small \citep[e.g.][]{zhang11} and hence they may not significantly affect the $g^0_3/g^0_1$ ratio. However, if they are somewhat comparable to the $Y^2_2$ amplitude (as suggested in some seismic studies) then the \citet{B2000} study suggests they would actually work to increase the $g^0_3/g^0_1$ ratio compared to a scenario with no $Y^0_2$ heat flux pattern. This could exacerbate the issue of matching the $g^0_3/g^0_1$ observations with dynamo models and would suggest that an even stronger or thicker stable layer may be needed. In addition, there are previously published dynamo models that do match the $\chi^2$ criterion and the $g^0_3/g^0_1$ ratio for particular parameter choices. For example, a previous study by \citet{D2008} 
captures the modern-day average $g^0_3/g^0_1$ ratio in a model with no stable layer, but in a low Rayleigh number regime with relatively large lateral CMB heat flux perturbations that may not be Earth-like. Furthermore, \citet{L2017} finds that the $g^0_3/g^0_1$ ratio may also depend on inner core size. That study implemented different buoyancy conditions relevant to various potential thermal histories for the inner core and showed that the $g^0_3/g^0_1$ ratio decreases as the inner core grows through time reaching values similar to present day observations for present day inner core sizes. These studies demonstrate that a stable layer may not be the only potential explanation for the present day $g^0_3/g^0_1$ ratio, but that other ingredients involving core thermal histories may contribute as well. \par



\section{Conclusion}
Our study demonstrates that a stably stratified layer at the top of Earth's core may be necessary to explain the zonal octupolar component of the geomagnetic field over the past 10,000 years. We found that a fairly thin stable layer (60 km) needs to be relatively strongly stratified ($(N/2\Omega)^2 \in [0.95, 1.22]$) whereas a moderately thick layer (130 km) needs to be more weakly stratified ($(N/2\Omega)^2 \in [0.14, 0.27]$). Our model with a 350 km thick stable layer could not match the geomagnetic zonal octupolar constraint and the Earth-like $\chi^2$ criterion.

\acknowledgments
  We thank W. Kuang and W. Jiang for their help with numerical dynamo model mMoSST. We also thank the anonymous reviewers for helpful comments that improved this manuscript. This research project was conducted using computational resources at the Maryland Advanced Research Computing Center (MARCC).  All numeric data from the numerical simulations used in the figures can be found in the supplementary information. The CALS10k.2 model data can be found at https://www.gfz-potsdam.de/en/section/geomagnetism/data-products-services/geomagnetic-field-models/.






%
%
%

\clearpage
 
 \clearpage

\listofchanges


\begin{thebibliography}{}
\bibitem[{\textit{Bloxham}}(2000)]{B2000} Bloxham, J. (2000). Sensitivity of the geomagnetic axial dipole to thermal core-mantle interactions. \textit{Nature.}
405, 63--65.
\bibitem[{\textit{Braginsky}}(1993)]{b1993} Braginsky, S. I. (1993). MAC-oscillations of the hidden ocean of the core. \textit{J. Geomagn. Geoelectr.}
45, 1517--1538.
\bibitem[{\textit{Braginsky}}(1999)]{b1999} Braginsky, S. I. (1999). Dynamics of the stably stratified ocean at the top of the core. \textit{Phys. Earth Planet. Int.}
111, 21--34.
\bibitem[{\textit{Bullard $\&$ Gellman}}(1954)]{B1954} Bullard, E., and Gellman, H. (1954). Homogenous dynamos and terrestrial magnetism. \textit{Phil. Trans. Roy. Soc. London. A.}
247, 213--278.
\bibitem[{\textit{Buffett}}(2014)]{B2014} Buffet, B. (2014). Geomagnetic fluctuations reveal stable stratification at the top of the Earth's core. \textit{Nature.}
507, 484--87. 
\bibitem[{\textit{Christensen et al.}}(2001)]{C2001} Christensen, U.R., Aubert, J., Cardin, P., Dormy, E., Gibbons, S., Glatzmaier, G., Grote, E., Honkura, Y., Jones, C., Kono, M., Matsushima, M., Sakuraba, A., Takahashi, F., Tilgner, A., Wicht, J., and Zhang, K. (2001). A numerical dynamo benchmark. \textit{Phys. Earth Planet. Inter.} 123, 25--34.
\bibitem[{\textit{Christensen $\&$ Wicht}}(2008)]{Ch2008} Christensen, U.R., and Wicht, J. (2008). Models of magnetic field generation in partly stable planetary cores: Applications to Mercury and Saturn. \textit{Icarus.}
196, 16--34.
\bibitem[{\textit{Christensen et al.}}(2010)]{C2010} Christensen, U. R., Aubert, J. and Gauthier, H. (2010). Conditions for Earth-like geodynamo models. \textit{Earth Planet. Sci. Lett.}
296, 487--96.
\bibitem[{\textit{Constable et al.,}}(2016)]{Con2016} Constable, C., Korte, M., and Panovska, S. (2016) Persistent high paleosecular variation activity in southern hemisphere for at least 10000 years. \textit{Earth Planet. Sci. Lett.}
453, 78--86.
\bibitem[{\textit{Davies et al.}}(2008)]{D2008} Davies, C., Gubbins, D., Willis, A. P., and Jimack, P. K (2008). Time-averaged paleomagnetic field and secular variation: Predictions from dynamo solutions based on lower mantle seismic tomography. \textit{Phys. Earth Planet. Sci}
169, 194--203.
\bibitem[{\textit{Davies $\&$ Constable}}(2014)]{D2014} Davies, C., and Constable, C. (2014). Insights from geodynamo simulations into long-term geomagnetic field behaviour. \textit{Earth Planet. Sci. Lett.} 404, 238--49.
\bibitem[{\textit{Gubbins}}(2007)]{G2007} Gubbins, D. (2007). Geomagnetic constraints on stratification at the top of Earth's core. \textit{Earth Planet. Space.}
59, 661--664.
\bibitem[{\textit{Gubbins $\&$ Davies}}(2012)]{G2012} Gubbins, D., and Davies, C. (2012). The stratified layer at the core-mantle boundary caused by barodiffusion of oxygen, sulphur and silicon. \textit{Earth Planet. Sci. Lett.}
215, 21--28.
\bibitem[{\textit{Gubbins et al.}}(2015)]{G2015} Gubbins, D., Alf$\acute{e}$, D., Davies, C., and Pozzo, M. (2015). On core convection and the geodynamo: Effects of high electrical and thermal conductivity. \textit{Phys. Earth Planet. Inter.}
247, 56--64.
\bibitem[{\textit{Heimpel and Evans}}(2013)]{He2013} Heimpel, M. H., and Evans, M. E. (2013). Testing the geomagnetic dipole and reversing dynamo models over Earth's cooling history. \textit{Phys. Earth Planet. Inter.}
224, 124--131.
\bibitem[{\textit{Helffrich}}(2013)]{H2013} Helffrich, G., and Kaneshima, S. (2013). Causes and consequences of outer core stratification. \textit{Earth Planet. Sci. Lett.}
223, 2--7.
\bibitem[{\textit{Jackson et al.}}(2000)]{J2000} Jackson, A., Jonkers, A. R. T and Walker, M. R. (2000). Four centuries of geomagnetic secular variation from historical records. \textit{Phil. Trans. R. Soc. Lond. A}
358, 957--990.
\bibitem[{\textit{Jiang $\&$ Kuang}}(2008)]{K2008} Jiang, W., and Kuang, W. (2008). An MPI-based MoSST core dynamics model. \textit{Phys. Earth Planet. Inter.}
170, 46--51.
\bibitem[{\textit{Kaneshima}}(2018)]{Ka2018} Kaneshima, S. (2018). Array analyses of SmKS waves and the stratification of Earth's outermost core. \textit{Phys. Earth Planet. Inter.}
276, 234--246.
\bibitem[{\textit{Kuang $\&$ Bloxham}}(1999)]{K99} Kuang, W., and Bloxham, J. (1999). Numerical Modeling of Magnetohydrodynamic Convection in a Rapidly Rotating Spherical Shell: Weak and Strong Field Dynamo Action. \textit{J. Comput. Phys.}
153, 51--81.

\bibitem[{\textit{Landeau et al.}}(2016)]{L2016} Landeau, M., Olson, P., Deguen, R., and Hirsh, B. H. (2016). Core merging and stratification following giant impact. \textit{Nature. Geosci.}
9, 786--789.
\bibitem[{\textit{Landeau et al.}}(2017)]{L2017} Landeau, M., Aubert, J., and Olson, P. (2017). The signature of inner-core nucleation on the geodynamo. \textit{Earth Planet. Sci. Lett}
465, 193--204.

\bibitem[{\textit{Lister $\&$ Buffett}}(1998)]{li1998} Lister, J. R., and Buffett, B. (1998). A stratification of the outer core at the core-mantle boundary. \textit{Phys. Earth Planet. Inter.}
105, 5--19.
\bibitem[{\textit{Olsen et al.}}(2014)]{O2014} Olsen, N., L$\ddot{u}$hr, H., Finlay, C. C., Sabaka, T. J., Michaelis, I., Rauberg, J., and Clausen, L. (2014). The CHAOS-4 geomagnetic field model. \textit{Geophy. J. Int.}
197, 815--827.
\bibitem[{\textit{Olson et al.}}(2017)]{O2017} Olson, P., Landeau, M., and Reynolds, E. (2017). Dynamo tests for stratification below the core-mantle boundary. \textit{Phys. Earth Planet. Inter.}
271, 1--18.
\bibitem[{\textit{Stanley $\&$ Bloxham}}(2004)]{S2004} Stanley, S., and Bloxham, J. (2004). Convective-region geometry as the cause of Uranus' and Neptune's unusual magnetic fields. \textit{Nature.}
428, 151--153.
\bibitem[{\textit{Stanley $\&$ Mohammadi}}(2008)]{S2008} Stanley, S., and Mohammadi, A. (2008). Effects of an outer thin stably stratified layer on planetary dynamos. \textit{Phys. Earth Planet. Inter.}
168, 179--90.
\bibitem[{\textit{Tanaka $\&$ Hamaguchi}}(1993)]{tanaka93} Tanaka, S., and Hamaguchi, H. (1993). Velocities and chemical stratification in the outermost core. \textit{J. Geomag. Geoelectr.}
45, 1287--1301.
\bibitem[{\textit{Tang et al.}}(2015)]{tang15} Tang, V., Zhao, L., and Hung, S. (2015). Seismological evidence for a non-monotonic velocity gradient in the topmost outer core. \textit{Sci. Rep}. 5, 8613.
\bibitem[{\textit{Zhang $\&$ Zhong}}(2011)]{zhang11} Zhang, N., and Zhong, S. (2011). Heat fluxes at the Earth's surface and core-mantle boundary since Pangea formation and their implications for the geomagnetic superchrons. \textit{Earth Planet. Sci. Lett.} 306, 205--216.
\end{thebibliography}
\end{document}


\supportinginfo{Sensitivity of the Geomagnetic Octupole to a stably stratified Layer in the Earth's Core}

 \authors{C. Yan\affil{1} and
 S. Stanley\affil{1,2}}

\affiliation{1}{Department of Earth and Planetary Science, Johns Hopkins University, MD 21218, USA.}
\affiliation{2}{Johns Hopkins University, Applied Physics Laboratory. Laurel, MD 20732, USA.}


\correspondingauthor{C. Yan}{cyan10@jhu.edu}

%
%

\section*{Contents}
\begin{enumerate}
\item Text S1
\item Figure S1 to S4
\item Table S1 to S2
\end{enumerate}


\section*{Numerical Model Details}
\subsection*{Fundamental equations and non-dimensionalization}
In the model, the Navier-Stokes, magnetic induction, and energy equations are non-dimensionalized using the same characteristic scales as in \citet{D2014}. The fundamental units are listed here for completeness of this study: the shell thickness of the dynamo region $d$ as a length scale, $\tau = d^2/\eta$ (where $\eta$ is the magnetic diffusivity) as a time scale, $\beta d$ (where $\beta$ is the buoyancy gradient at the outer core boundary) as a buoyancy scale and $\sqrt{2\rho\Omega\mu_0\eta}$ (where $\Omega$ is the angular velocity, $\rho$ is the density and $\mu_0$ is the permeability of vacuum) is the magnetic field intensity scale. The non-dimensional equations are: 
 \begin{equation}
\frac{E}{P_m}(\partial_t + \mathbf{v} \cdot \nabla)\mathbf{v} + \mathbf{l}_z \times \mathbf{v} = -\nabla p + \mathbf{J}\times \mathbf{B} + R_{a} \Theta \mathbf{r} + E \nabla^2 \mathbf{v} 
\end{equation}
\begin{equation}
 (\partial_{t} - \nabla^2) \mathbf{B} = \nabla \times (\mathbf{v} \times \mathbf{B}) 
 \end{equation}
\begin{equation}
(\partial_{t} - \frac{P_m}{P_r} \nabla^2) \Theta = -\mathbf{v} \cdot \nabla[C_0(r) + \Theta] + Q
 \end{equation}
where $\mathbf{v}$ and $\mathbf{B}$ are the velocity and magnetic perturbation fields. $\Theta$ is the buoyancy perturbation field from the static co-density state $C_0$. The total co-density field $C$ is therefore written as $C = C_0 + \Theta$ (further information on this variable in the next section). $Q$ represents the internal buoyancy sources and $p$ is the modified pressure, $\mathbf{J}$ the current density and $\mathbf{l_z}$ is the unit vector in the direction of the rotation axis. The non-dimensional parameters in the equations are the Ekman number, the modified Rayleigh number, the Prandtl number and magnetic Prandtl number which are defined by: 
\begin{equation}
E \equiv \frac{\nu}{2\Omega d^2}
 \end{equation}
 \begin{equation}
P_r \equiv \frac{\nu}{\kappa}
 \end{equation}
  \begin{equation}
P_m \equiv \frac{\nu}{\eta}
 \end{equation}
  \begin{equation}
R_a \equiv \frac{\alpha g_0 \beta d^4}{\nu\kappa} 
 \end{equation}
where $\nu$ is the kinematic viscosity, $\kappa$ the buoyancy diffusivity, $\alpha$ the buoyancy expansion coefficient and $g_0$ is the gravitational acceleration. 

\subsection*{Implementation of stratification}

In our simulations, we use the buoyancy variable $C$, termed the "co-density", which combines the buoyancy effects from the perturbations from the core's mean state of the light-element  concentration $\Delta \chi_c$ (not to be confused with Earth-like criterion $\chi^2$) and the temperature $\Delta T$. The non-dimensional co-density is given by 
 \begin{equation}
  C = \alpha\Delta T + \alpha_c \Delta \chi_c
 \end{equation}
where $\alpha$ and $\alpha_c$ are thermal and compositional expansion coefficients, respectively.
The gradient of the co-density $\nabla C$ thus measures the buoyancy flux. In the convecting region, the reference state co-density gradient can be written
  \begin{equation}
\frac{d C_0(r)}{dr} = \frac{\beta^{(b)}}{r^2} + \beta^{(i)} r + \beta^{(s)} r^3
 \end{equation}
 where the constants $\beta^{(b)}, \beta^{(i)}$ and $\beta^{(s)}$ define the amount of the bottom, internal and non-uniform specific buoyancy sources separately. We use the same parameter values as model C1-4* of \citet{D2014}, where the bottom buoyancy constant $\beta^{(b)} = 9.33$ and the internal buoyancy constant $\beta^{(i)} = -1.67$, whereas the specific buoyancy source is small enough to be neglected \citep{D2011}.  
We set a specific radius for the onset of the stably stratified layer where the reference co-density gradient in equation (9) above is replaced by a prescribed positive constant A,
  \begin{equation}
 \frac{dC_0(r)}{dr} = A
 \end{equation}
 that we vary to adjust the strength of the stratification.
 The parameter A can be related to the $Brunt-V\ddot{a}is\ddot{a}l\ddot{a}$ frequency $N$ \citep{G1982}, which is given by
    \begin{equation}
 N^2 = -g[\frac{\rho g}{\kappa_s} + \frac{1}{\rho}\frac{d\rho}{dr}] \approx -\frac{g}{\rho}\frac{d\rho}{dr}
 \end{equation}
  where $\kappa_s$ is the bulk modulus. Keeping the dominant term we can relate the strength of the stratification $N/2\Omega$ to the parameter A by equation
   \begin{equation}
(\frac{N}{2\Omega})^2 \approx \frac{R_a E^2}{P_r} A
 \end{equation}



\section*{Figure S1}
Same as Figure 2 but for the zonal quadrupole to dipole ratio. 
\begin{figure}[h]
\centering
\includegraphics[width=25pc]{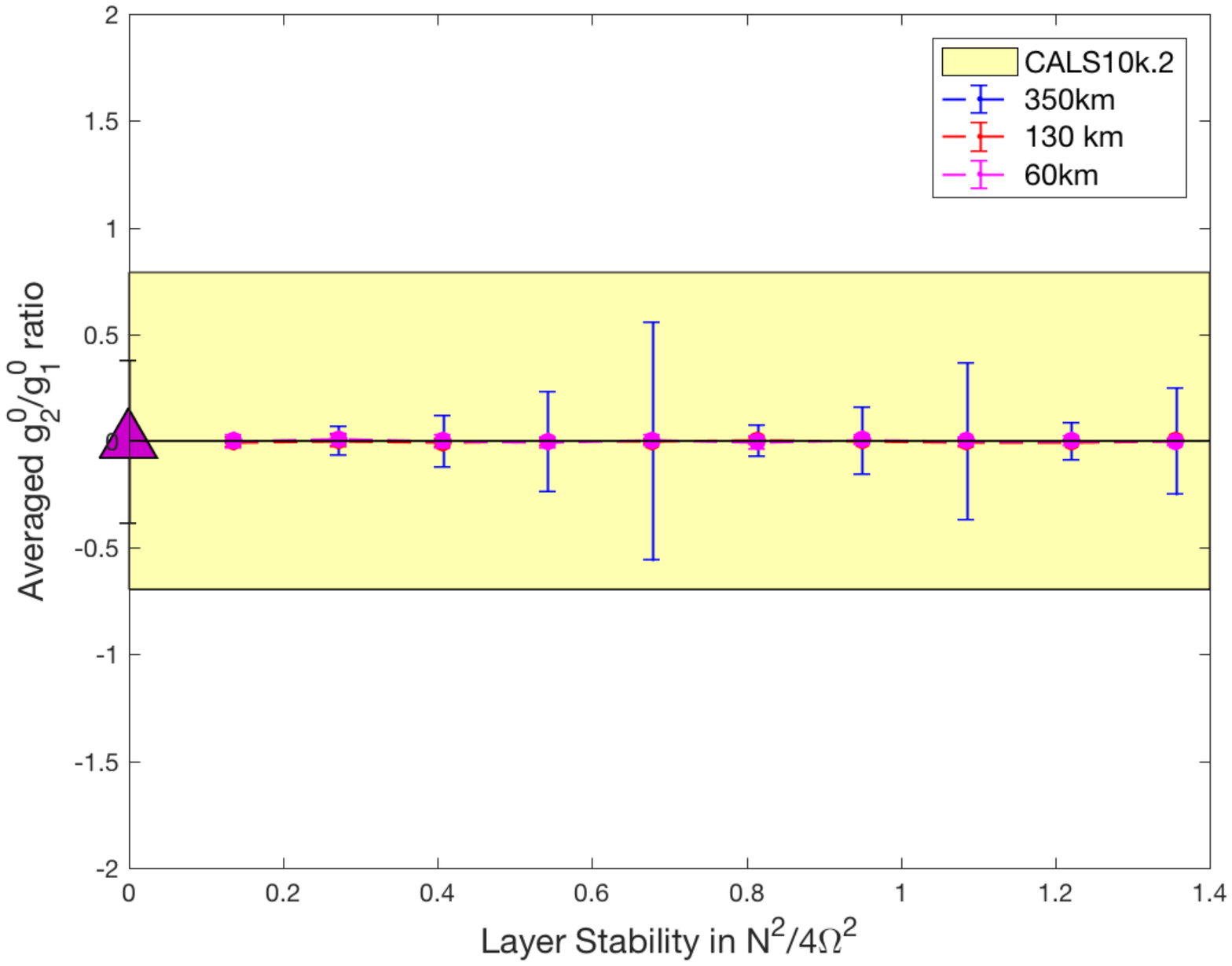}
\vspace{-1cm}
\label{figone}
\end{figure}

\section*{Figure S2}
Same as Figure 3 but for the zonal quadrupole to dipole ratio. 
\begin{figure}[h]
\centering
\includegraphics[width=25pc]{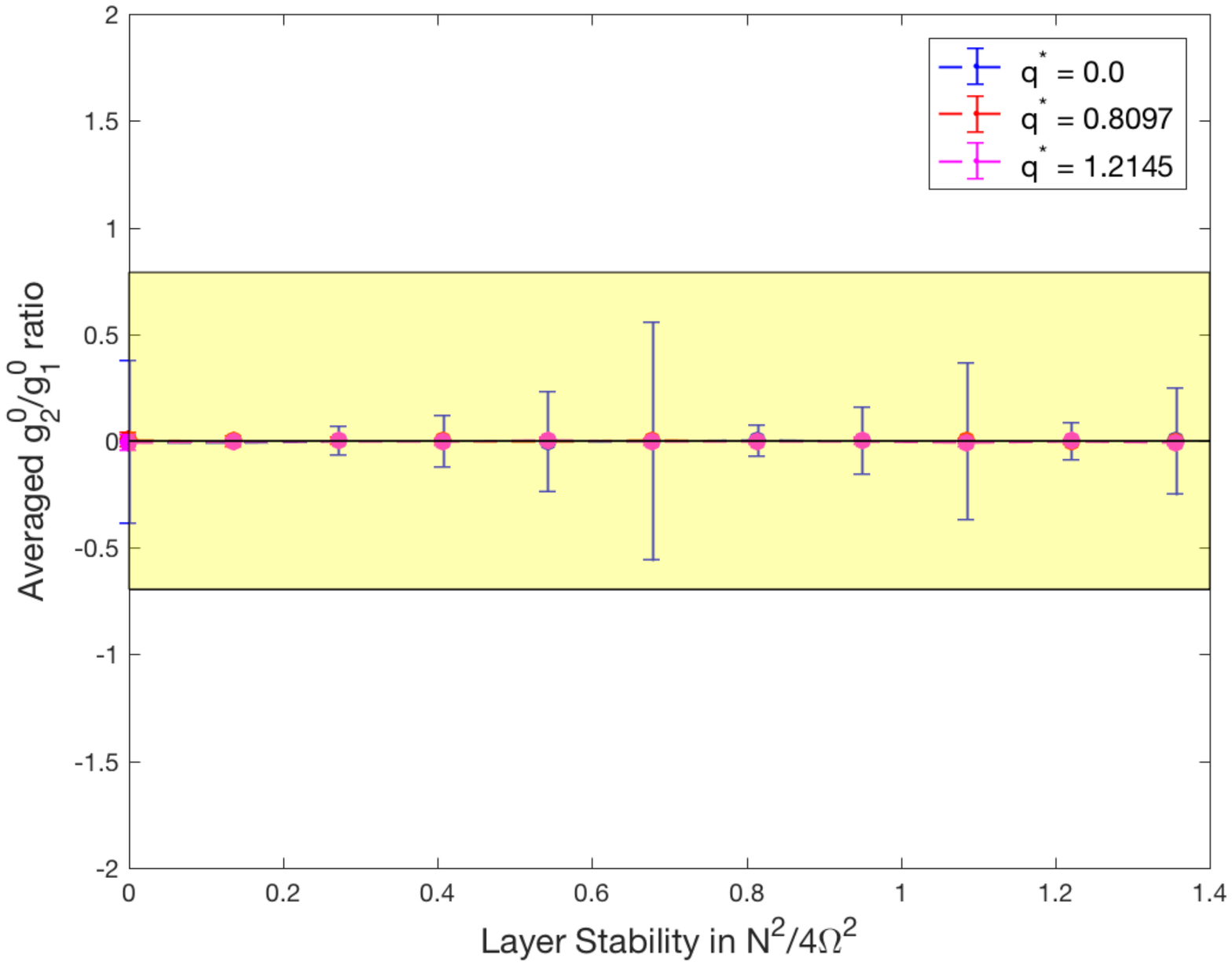}
\label{figone}
\end{figure}

\section*{Figure S3}
(a) The non-dimensional
radial velocity $\mathbf{v'}$ profiles in models with a 350 km stable layer and various layer stratification $(N/2\Omega)^2$, where $\mathbf{v'} = \mathbf{v} d/\eta$ for a dimensional velocity $\mathbf{v}$. Note the magnetic Reynolds number in our simulations range from 110 to 150. The x axis shows the outer core region from ICB (1225 km) to CMB (3500 km) whereas the y axis shows the magnitude of the radial velocity averaged over 1 magnetic diffusion time and the spherical surface at each radius. The black dash-dot line marks the boundary of the implemented stable layer (i.e. $r_{CMB} - \Delta r$). (b) The time averaged zonal flows in a model without a stable layer (left column) and with a 350 km stable layer and $(\frac{N}{2\Omega})^2 = 1.3564$ (right column). 
\begin{figure}[h]
\centering
\includegraphics[width=40pc]{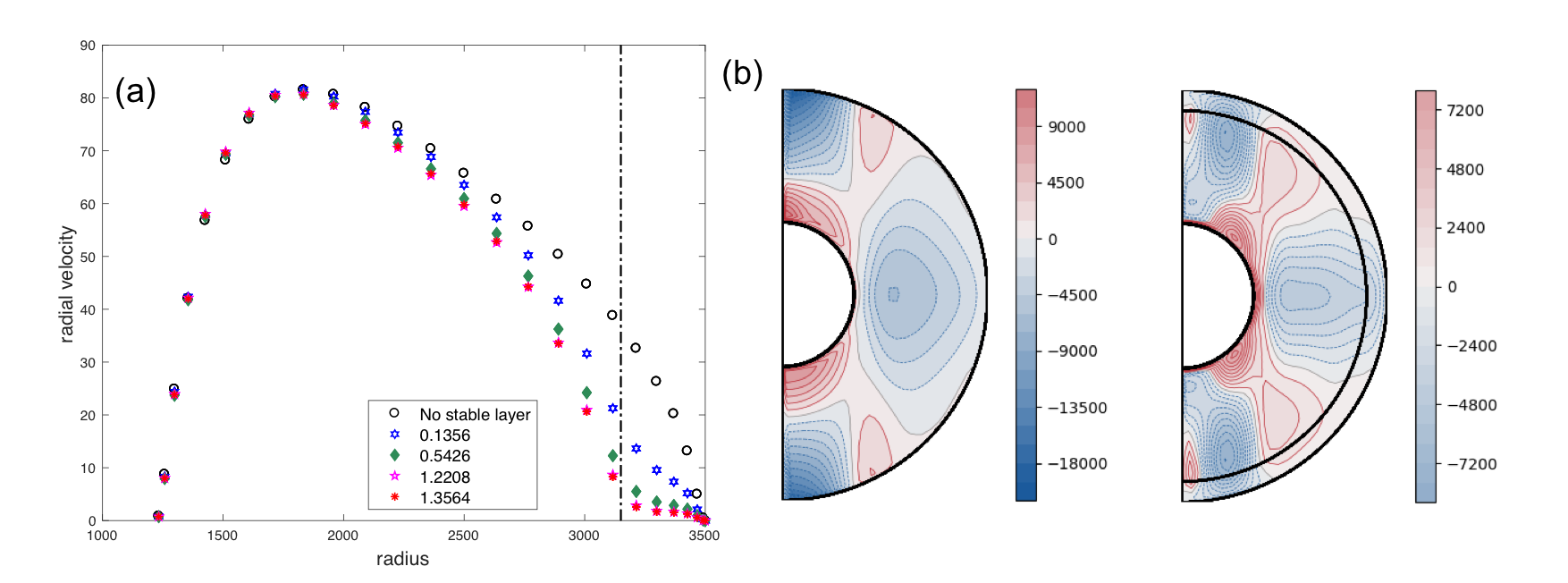}
\label{figone}
\end{figure}

\section*{Figure S4}
(a) Snapshots of the radial magnetic field at the CMB and (b) time-averaged axisymmetric buoyancy wind profiles in a model without a stable layer (left column) and a 130 km stable layer model with $(N/2\Omega)^2 = 0.2713$ (right column).
\begin{figure}[h]
\centering
\includegraphics[width=35pc]{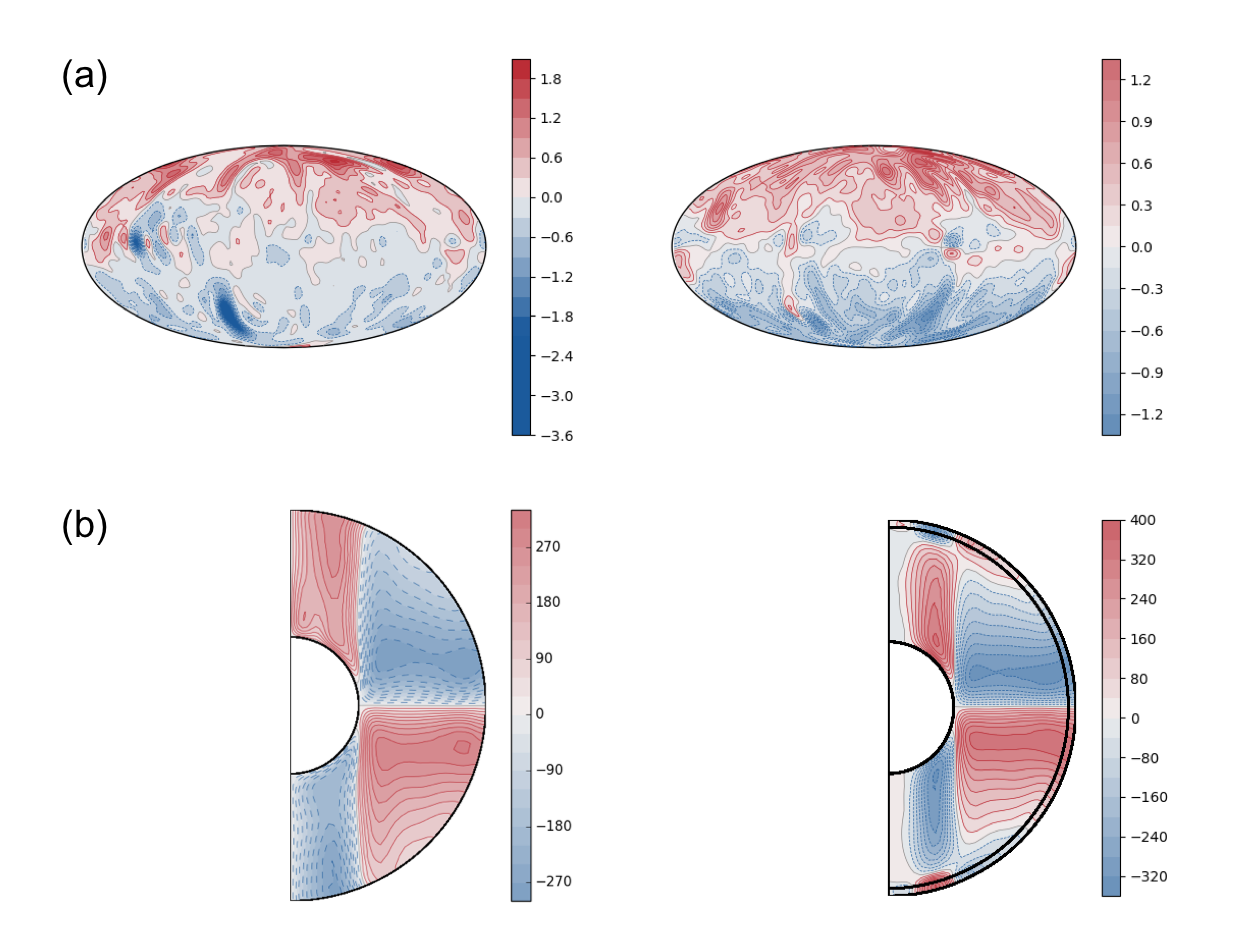}
\label{figone}
\end{figure}


 
 \section*{Table S1}
 $\chi^2$ evaluation for our models in Table 1 as well as the models with core-mantle boundary heat flux variations. The underlined cases are in good agreement with the geomagnetic observations based on the $\chi^2$ criterion. The bold cases match both the $\chi^2$ criterion as well as the $g^0_3/g^0_1$ ratio constraint. The definition of $q^*$ is in the Figure 3 caption.\par
\begin{sidewaystable}[h]
 \hspace{-3.5cm}
 \centering
 \begin{tabular}{lcccccccccccc}
 \hline
 \multicolumn{12}{ c }{layer stratification $(N/2\Omega)^2$} \\
 \hline
    $\Delta r (km)$ , $q^*$ & 0 & 0.1356 & 0.2713 & 0.4069 & 0.5426 & 0.6782 & 0.8138 & 0.9495 & 1.0851  &  1.2208   & 1.3564 \\
 \hline
   60 km & \underline{1.0851} & \underline{0.6499} & \underline{0.6270} & \underline{1.4019} & \underline{1.2453} & \underline{1.9710} & \underline{2.2532} & $\mathbf{2.8439}$ & $\mathbf{3.5152}$ & $\mathbf{3.7153}$ & 4.0938 \\  \hline
  130 km & \underline{0.3676} & $\mathbf{1.8701}$ & $\mathbf{3.6273}$ & 4.8262   & 5.7901   & 6.6318 & 7.0833  & 7.5562 & 7.5545 & 8.0796  & 8.2228 \\
 \hline
   350 km & \underline{0.5934} & 7.1184   & 10.347   & 12.565   & 14.421   & 15.848   & 16.975   & 17.395   & 16.783 & 18.379   & 18.495\\
 \hline
 350 km,  & \underline{1.3112} & 5.9553    & 7.6981   & 9.3450     & 10.679   & 12.328    & 12.265   & 10.143   & 12.853   &14.276   & 15.110 \\
  $q^* = 0.8097$\\
  \hline
   350 km, & \underline{1.6347}    & 6.7317    & 7.5691    & 8.3400    & 9.5470   & 10.512   & 10.586   & 11.875   & 12.092   & 12.462   & 12.672\\
  $q^* = 1.2145$\\ 
  \hline
\end{tabular}
\end{sidewaystable}

\section*{Table S2}
Averaged values of the $g^0_3/g^0_1$ ratio with the standard deviation for our models in Table 1 as well as the models with core-mantle boundary heat flux variations. \par
\begin{sidewaystable}
 \hspace{50cm}
 \centering
 \begin{tabular}{lcccccccccccc}
 \hline
 \multicolumn{12}{ c }{layer stratification $(N/2\Omega)^2$} \\
 \hline
    $\Delta r (km)$ , $q^*$ & 0 & 0.1356 & 0.2713 & 0.4069 & 0.5426 & 0.6782 & 0.8138 & 0.9495 & 1.0851  &  1.2208   & 1.3564 \\
 \hline
   60 km & 0.0409 & 0.0335  & 0.0208 & 0.0176  & 0.0191 & 0.0122  & 0.0148 & 0.0022  & 0.0008  & -0.0002  & -0.0032\\  
    & $\pm 0.0193$ & $\pm 0.0170$ &  $\pm 0.0145$ & $\pm 0.0184$ & $ \pm 0.0178$ & $\pm 0.0126$ & $ \pm 0.0132$ & $\pm 0.0094$ & $\pm 0.0076$ & $\mp 0.0281$ & $ \mp 0.0088$ \\
   \hline
  130 km & 0.0394 & 0.0109 & 0.0025 & -0.0042   & -0.0084 & -0.0082 & -0.0134  & -0.0140 & -0.0166 & -0.0167  & -0.0192 \\
  & $\pm 0.0171$ & $\pm 0.0132$ &  $\pm 0.0095$ & $\mp 0.0103$ & $ \mp -0.0081$ & $\mp 0.0084$ & $ \mp 0.0079$ & $\mp 0.0084$ & $\mp 0.0075$ & $\mp 0.0070$ & $ \mp 0.0074$ \\
 \hline
   350 km & 0.0401 & -0.0013   & -0.0121   & -0.0172   & -0.0183   & -0.0234   & -0.0199   & -0.0216   & -0.0239 & -0.0196   & -0.0195\\
   & $\pm 0.0170$ & $\mp 0.0099$ & $\mp 0.0064$ & $\mp 0.0072$ & $\mp 0.0050$ & $\mp 0.0056$ & $\mp 0.0063$ & $\mp 0.0066$ & $\mp 0.0058$ & $\mp 0.0058$ & $\mp 0.0060 $ \\
 \hline
 350 km,  & 0.0470  & 0.0074 & -0.0065 & -0.0133 & -0.0169 & -0.0186 &  -0.0195 & -0.0221 & -0.0222 & -0.0208 & -0.0211\\
  $q^* = 0.8097$ & $\pm 0.0220$ & $\pm 0.0109$ & $\mp 0.0107$ & $\mp 0.0067$ & $\mp 0.0072$ & $\mp 0.0059$ & $\mp 0.0063$ & $\mp 0.0064$ & $\mp 0.0057$ & $\mp 0.0068$ & $\mp 0.0073$ \\
  \hline
   350 km, & 0.0433 & 0.0155 & -0.0014 & -0.0144 & -0.0185 & -0.0204 & -0.0212 & -0.0221 & -0.0220 & -0.0225 & -0.0233 \\
  $q^* = 1.2145$ & $\pm 0.0248$ & $\pm 0.0151$ & $\mp 0.0117$ & $\mp 0.0094$ & $\mp 0.0098$ & $\mp 0.0084$ & $\mp 0.0077$ & $\mp 0.0079$ & $\mp 0.0077$ & $\mp 0.0077$ & $\mp 0.0080$\\ 
  \hline
\end{tabular}
\end{sidewaystable}









